# AN EFFICIENT PARALLEL ALGORITHM FOR COMPUTING DETERMINANT OF NON-SQUARE MATRICES BASED ON RADIC'S DEFINITION

Neda Abdollahi[1], Mohammad Jafari[1], Morteza Bayat[2], Ali Amiri[3], Mahmood Fathy[4]

[1]Department of Electronic and Computer Engineering, Zanjan Branch, Islamic Azad University, Zanjan, Iran
[2]Department of Mathematics, Zanjan Branch, Islamic Azad University, Zanjan, Iran
[3]Computer Engineering Group, University of Zanjan, Zanjan, Iran
[4]Department of Computer Engineering, Iran University of Science and Technology, Tehran, Iran

## ABSTRACT

*One of the most significant challenges in Computing Determinant of Rectangular Matrices is high time complexity of its algorithm. Among all definitions of determinant of rectangular matrices, Radic's definition has special features which make it more notable. But in this definition, $C\binom{N}{M}$ sub matrices of the order $m \times m$ needed to be generated that put this problem in np-hard class. On the other hand, any row or column reduction operation may hardly lead to diminish the volume of calculation. Therefore, in this paper we try to present the parallel algorithm which can decrease the time complexity of computing the determinant of non-square matrices to $O(N^2)$.*

## 1. INTRODUCTION

Determinant is one of the basic concepts in linear algebra and applied statistics that have major applications in various branches of mathematics and engineering. Computing the determinant of a matrix is a classical problem, which is addressed in normal forms of matrix studies [1-4] and computational number theory [5].

In principle, determinant is only defined for square matrices [6]. There is usable and clear definition for the calculation of square matrices determinant. A parallel algorithm for the calculation of $n \times n$ square matrix determinant is presented with time complexity $\theta(n)$ [7]. But, extracting data from physical phenomena and real world applications generally leads to produce non-square matrices [8-10].

So far, many definitions for determinant of non-square matrices are given. Most of the works that has been done, focusing on the definition and calculation the determinant of non-square matrices by dividing them into square blocks [11][12] .[13]. In reference [12] Radic proposed an efficient definition for determinant of non-square matrices that has most of the important properties of square matrices determinant. Also, some other properties of Radic's determinant and its geometrical interpretations, involving polygons in the plan R2 and polyhedral in R3 are given in[12][14-18].

In [19], non-square matrices are converted to square matrices by summarizing, that leads to miss some part of data.

The determinant of non-square matrix is used in retrieving images with different sizes [8]. Also, there are some works on video retrieval and video shot boundary detection and image processing by using determinant of non-square matrix [8][20-23]. Thus providing an effective solution for calculating the determinant of non-square matrices can be very valuable and helpful.

In paper [24] a parallel algorithm based on pointer jumping technique is proposed to calculate the determinant of non-square matrices of order $2 \times n$. But despite the successful work that has been done for the definition of non-square matrices determinant, yet there isn't any efficient algorithm to compute this determinant.

According to Radic's definition for the determinant of a non-square matrix m × n, it should be calculated the determinant of $C_{\binom{n}{m}}$ square matrices from the order of m × m. The square matrices are obtained by combination of non-square matrix columns. Hence, the calculation of non-matrices determinant is NP-hard. Some researchers [14] have tried decrease rows or columns of a non-square matrix to convert it into a square matrix. But normally any change in rows or columns of non-square matrices increase computation and column operations.

According to above explanation and proven theorems [12] currently, the only effective solution to compute the determinant of non-square matrices by acceptable time complexity is parallel algorithms.

To paralyze the algorithm, at first, the dependency between each Radic's sub-square matrices should be omitted. Secondly, each of these determinants also needs to be computed in parallel.

In this paper, we proposed a parallel algorithm to calculate the determinant of non-square matrices based on Radic's definition with $O(n^2)$ time complexity.

Problems and motivations are considered in Section 2. In Section 3 the Radic's definition are analyzed in details. Section 4 includes the proposed method to compute each arbitrary elements of Dictionary order independently and in Section 5 a parallel algorithm for computing Raidc's determinant is presented. The complexity of proposed algorithm is perused due to the hardware architecture In Section 6. Section 7 clarifies our conclusions.

## 2. PROBLEMS AND MOTIVATIONS

Radic's definition [12] for calculating the determinant of non-square matrices has numerous significant properties and advantages in comparing to other definitions. Specially, it has almost all the properties of determinant of square matrices [12].

According to Radic's definition, it is evident that the determinant of a non-square matrix can be computed as sum of specially signed square sub matrices. These sub matrices is obtained by calculating specific permutation of columns of non-square matrix. Although this definition is easy to compute and understand, it has exponential time complexity. In other words, computing the $\det(A)$ requires to compute determinants of $C\binom{n}{m}$ square sub matrices of order $m \times m$, which lead to exponential time complexity. Regarding to the previous works, it is obvious that applying column and row operations for computing the determinant of non-square matrices is inefficient [25]. In addition, due to the dependency between determining square sub matrices, it is impossible to design an efficient parallel algorithm based on this definition.

In this paper we propose a novel approach for parallel production of square sub matrices which reduces the time complexity to $O(m \times (n - m))$.

# 3. DETAILED ANALYSIS OF RADIC'S DEFINITION

At first, we will clarify some preliminary concepts and then assess Radic's definition due to these concepts.

**Definition 1: ascending sequence**

A sequence of elements of a partially ordered set such that each member of the sequence is less than the following one. So, for set $A = \{1, 2, 3, \ldots, n\}$, each sub set $B = \{a_1, a_2, \ldots, a_m\}$ is an ascending sequence if condition $\forall (a_1, a_2, \ldots, a_m) \in A$ and $(m < n)$ and $(a_1 < a_2 < \cdots < a_m)$ is satisfied.

**Definition 2: dictionary order**

Suppose $\{A_1, A_2, \ldots, A_n\}$ is an n-tuple of sets, with respective total orderings $\{<_1, <_2, \ldots, <_n\}$. The dictionary ordering $<^d$ of $A_1 \times A_2 \times \ldots \times A_n$ is then

$(a_1, a_2, \ldots, a_n) <^d (b_1, b_2, \ldots, b_n) \Leftrightarrow (\exists m > 0)(\forall i \leq m)(a_i = b_i) \wedge (a_m <_m b_m)$. That is, if one of the terms $a_m <_m b_m$ and all the preceding terms are equal.

**Theorem 1:**

Regarding to def.1, the maximum number of m-tuple sub sequences of ascendant set $A = \{a_1, a_2, \ldots, a_n\}$ where $m < n$, is equal to $\binom{n}{m}$.

**Proof:**

Putting the minimum element in the first place (as $a_1$ in set A), we would have $n - 1$ choice for the remaining $m - 1$ places. In the same way, by selecting $a_2$, there will be $n - 2$ selection for the last $m - 1$ places, and finally by putting $a_{n-m+1}$ in the first place, there will be remained $m - 1$ choices for $m - 1$ places. In other word, all the possible selections can be shown as follow.

$$
\begin{array}{ll}
\textbf{1:} & \underbrace{a_1}_{p_1}, \underbrace{\overset{\binom{n-1}{m-1}}{p_2, \ldots, p_m}} \\[2em]
\textbf{2:} & \underbrace{a_2}_{p_1}, \underbrace{\overset{\binom{n-2}{m-1}}{p_2, \ldots, p_m}} \qquad A=\{a_1, a_2, \ldots, \underbrace{a_{n-m+1}, \ldots, a_n}_{m}\} \\[2em]
\cdots & \cdots \\[2em]
\textbf{n-m+1:} & \underbrace{a_{n-m+1}}_{p_1}, \underbrace{\overset{\binom{m-1}{m-1}}{p_2, \ldots, p_m}}
\end{array}
$$

In this case, all ascending sequences that can be produced is equal

$$\binom{n-1}{m-1} + \binom{n-2}{m-1} + \cdots + \binom{m}{m-1} + \binom{m-1}{m-1} = \binom{n}{m}. \blacksquare$$

According to the dictionary order and ascending sequences definition, it's obvious that, for $m < n$, the first element of $A = \{1,2,\dots,n\}$ is $[1,2,\dots,m]$, which we entitled *First Member*. Also, the last element in this sequence will be $[n-m+1, n-m+2, \dots, n]$ and the remaining sub ascending sequences will be in this interval.

$$[1,2,\dots,m] < [1,2,\dots,m+1] < \cdots < [n-m+1, n-m+2, \dots, n]$$

Now, according to Theorem 1, the sequences can be numbered from 0 to $\binom{n}{m} - 1$. Also, according to the latest member of ascending sequences the maximum value of each place is determined. For example, the maximum value which the $m^{\text{th}}$ place can be obtained, is n, But due to the need to establish the condition $a_{m-1} < a_m$, the value of $(m-1)^{\text{th}}$ place cannot exceed n-1.

In the following, we present Radic's definition for determinant of non-square matrices.

**Definition 3.** Let $A = [a_{i,j}]$ be an $m \times n$ matrix with $m \leq n$. The determinant of $A$, is defined as:

$$\det(A) = \sum_{1 \leq j_1 < \cdots < j_m \leq n} (-1)^{r+s} det \begin{bmatrix} a_{1j_1} & \cdots & a_{1j_m} \\ \vdots & \ddots & \vdots \\ a_{mj_1} & \cdots & a_{mj_m} \end{bmatrix}, \tag{1}$$

where $j_1, j_2, \dots, j_m \in \mathbb{N}, r = 1 + 2 + \cdots + m$ and $s = j_1 + \cdots + j_m$. If $m > n$, then we define $\det(A) = 0$.

Now, according to Definition 3 is observed that the following sub square matrices produced in Radic's definition is in accordance with the dictionary order. So, if an efficient algorithm can be represented for the computation of dictionary order elements, therefore Radic's determinant can also be calculated with greater efficiency.

## 4. COMPUTATION OF DICTIONARY SEQUENCE ELEMENTS

In this section, we attempt to compute each arbitrary elements of Dictionary order independently. In other word, by giving a q where $0 \leq q < \binom{n}{m}$, we try to calculate the $q^{\text{th}}$ element in the sequence. Due to this purpose, a novel definition entitled *combinatorial addition* is presented and table 1 is formed by the elements of Pascal's triangle.

| j \ i | i = 1 | i = 2 | … | i = n − m − 1 | i = n − m |
|---|---|---|---|---|---|
| j = 0 | $\binom{1}{0}$ | $\binom{2}{0}$ | … | $\binom{n-m-1}{0}$ | $\binom{n-m}{0}$ |
| j = 1 | $\binom{2}{1}$ | $\binom{3}{1}$ | … | $\binom{n-m-2}{1}$ | $\binom{n-m-1}{1}$ |
| … | … | … | … | … | … |
| j = m − 2 | $\binom{m-1}{m-2}$ | $\binom{m}{m-2}$ | … | $\binom{n-3}{m-2}$ | $\binom{n-2}{m-2}$ |
| j = m − 1 | $\binom{m}{m-1}$ | $\binom{m+1}{m-1}$ | … | $\binom{n-2}{m-1}$ | $\binom{n-1}{m-1}$ |

Table1 : Pascal's triangle

As you can see in Table 1, each (j, i)$^{\text{th}}$ entries in the table is obtained from $\binom{i+j}{j}$. According to Table 1and as it mentioned in theorem 1, the weight of each element in the Ascending sequence is equal the last column of the table.

$$\underset{1}{\binom{n-1}{m-1}}, \underset{2}{\binom{n-2}{m-2}}, \underset{3}{\binom{n-3}{m-3}}, \dots, \underset{k-1}{\binom{n-k+1}{m-k+1}}, \underset{k}{\binom{n-k}{m-k}}, \dots, \underset{m}{\binom{n-m}{0}}$$

Now, if $\binom{n-k}{m-k} < q \leq \binom{n-k+1}{m-k+1}$, $m-k+1$ element of the First Member will change. We use Table 1 to calculate the amount of the change. To this, according Table 1, we must go to left in the $j^{th}$ row where $\binom{n-k}{m-k}$ is located. As can be seen in Table 1, the first element at the left side of the start point is $\binom{n-k-1}{m-k-1}$.

Until condition $q \geq \binom{n-k}{m-k} + \dots + \binom{n-k-p}{m-k}$ is satisfied, we continue the steps to the left.

Then the numbers of steps, which is moved to the left side in $j^{th}$ row, will be added to the value of last m-k locations in the First Member. The new value of $q$ is calculated from the following equation.

$$q \leftarrow q - \sum_{i=0}^{p}\binom{n-k-i}{m-k}.$$

Then until $q = 0$, the algorithm is continued from $(n-m-p)^{th}$ column by the new value of $q$.

Example 1: for set $A = \{1, 2, \dots, 8\}$, the five-member ascending sequences in dictionary order is shown in table 2.

<div align="center">

N=8       $\binom{8}{5} = 56$

M=5

</div>

| $B_0$ | 1 | 2 | 3 | 4 | 5 | $B_{11}$ | 1 | 2 | 4 | 5 | 7 | $B_{22}$ | 1 | 3 | 4 | 5 | 8 | $B_{33}$ | 1 | 4 | 6 | 7 | 8 | $B_{44}$ | 2 | 3 | 6 | 7 | 8 |
|---|---|---|---|---|---|---|---|---|---|---|---|---|---|---|---|---|---|---|---|---|---|---|---|---|---|---|---|---|---|
| $B_1$ | 1 | 2 | 3 | 4 | 6 | $B_{12}$ | 1 | 2 | 4 | 5 | 8 | $B_{23}$ | 1 | 3 | 4 | 6 | 7 | $B_{34}$ | 1 | 5 | 6 | 7 | 8 | $B_{45}$ | 2 | 4 | 5 | 6 | 7 |
| $B_2$ | 1 | 2 | 3 | 4 | 7 | $B_{13}$ | 1 | 2 | 4 | 6 | 7 | $B_{24}$ | 1 | 3 | 4 | 6 | 8 | $B_{35}$ | 2 | 3 | 4 | 5 | 6 | $B_{46}$ | 2 | 4 | 5 | 6 | 8 |
| $B_3$ | 1 | 2 | 3 | 4 | 8 | $B_{14}$ | 1 | 2 | 4 | 6 | 8 | $B_{25}$ | 1 | 3 | 4 | 7 | 8 | $B_{36}$ | 2 | 3 | 4 | 5 | 7 | $B_{47}$ | 2 | 4 | 5 | 7 | 8 |
| $B_4$ | 1 | 2 | 3 | 5 | 6 | $B_{15}$ | 1 | 2 | 4 | 7 | 8 | $B_{26}$ | 1 | 3 | 5 | 6 | 7 | $B_{37}$ | 2 | 3 | 4 | 5 | 8 | $B_{48}$ | 2 | 4 | 6 | 7 | 8 |
| $B_5$ | 1 | 2 | 3 | 5 | 7 | $B_{16}$ | 1 | 2 | 5 | 6 | 7 | $B_{27}$ | 1 | 3 | 5 | 6 | 8 | $B_{38}$ | 2 | 3 | 4 | 6 | 7 | $B_{49}$ | 2 | 5 | 6 | 7 | 8 |
| $B_6$ | 1 | 2 | 3 | 5 | 8 | $B_{17}$ | 1 | 2 | 5 | 6 | 8 | $B_{28}$ | 1 | 3 | 5 | 7 | 8 | $B_{39}$ | 2 | 3 | 4 | 6 | 8 | $B_{50}$ | 3 | 4 | 5 | 6 | 7 |
| $B_7$ | 1 | 2 | 3 | 6 | 7 | $B_{18}$ | 1 | 2 | 5 | 7 | 8 | $B_{29}$ | 1 | 3 | 6 | 7 | 8 | $B_{40}$ | 2 | 3 | 4 | 7 | 8 | $B_{51}$ | 3 | 4 | 5 | 6 | 8 |
| $B_8$ | 1 | 2 | 3 | 6 | 8 | $B_{19}$ | 1 | 2 | 6 | 7 | 8 | $B_{30}$ | 1 | 4 | 5 | 6 | 7 | $B_{41}$ | 2 | 3 | 5 | 6 | 7 | $B_{52}$ | 3 | 4 | 5 | 7 | 8 |
| $B_9$ | 1 | 2 | 3 | 7 | 8 | $B_{20}$ | 1 | 3 | 4 | 5 | 6 | $B_{31}$ | 1 | 4 | 5 | 6 | 8 | $B_{42}$ | 2 | 3 | 5 | 6 | 8 | $B_{53}$ | 3 | 4 | 6 | 7 | 8 |
| $B_{10}$ | 1 | 2 | 4 | 5 | 6 | $B_{21}$ | 1 | 3 | 4 | 5 | 7 | $B_{32}$ | 1 | 4 | 5 | 7 | 8 | $B_{43}$ | 2 | 3 | 5 | 7 | 8 | $B_{54}$ | 3 | 5 | 6 | 7 | 8 |
|  |  |  |  |  |  |  |  |  |  |  |  |  |  |  |  |  |  |  |  |  |  |  |  | $B_{55}$ | 4 | 5 | 6 | 7 | 8 |

Table 2: all five-member subsets

For m = 5 and n = 8, we have table 3.

|  | $i = 0$ | $i = 1$ | $i = 2$ | $i = 3$ |
|---|---|---|---|---|
| $j = 0$ | $\binom{0}{0}$ | $\binom{1}{0}$ | $\binom{2}{0}$ | $\binom{3}{0}$ |
| $j = 1$ | $\binom{1}{1}$ | $\binom{2}{1}$ | $\binom{3}{1}$ | $\binom{4}{1}$ |
| $j = 2$ | $\binom{2}{2}$ | $\binom{3}{2}$ | $\binom{4}{2}$ | $\binom{5}{2}$ |
| $j = 3$ | $\binom{3}{3}$ | $\binom{4}{3}$ | $\binom{5}{3}$ | $\binom{6}{3}$ |
| $j = 4$ | $\binom{4}{4}$ | $\binom{5}{4}$ | $\binom{6}{4}$ | $\binom{7}{4}$ |

$\binom{i+j}{j}$   $\approx$

| 1 | 1 | 1 | 1 |
|---|---|---|---|
| 1 | 2 | 3 | 4 |
| 1 | 3 | 6 | 10 |
| 1 | 4 | 10 | 20 |
| 1 | 5 | 15 | 35 |

Table 3a                     table 3b

We assume q = 49. In this case, according to table 3, the weight of each place in the First Member would be as follows.

$$\binom{7}{4}\binom{6}{3}\binom{5}{2}\binom{4}{1}\binom{3}{0}$$
$$1 \quad 2 \quad 3 \quad 4 \quad 5$$

Since $\binom{7}{4} < q < \binom{8}{5}$, in the fifth row (j = 4) we will proceed to the left, but because $\binom{6}{4} + \binom{7}{4} > 49$ is not acceptable, so we stopped at p = 1. Therefore, p = 1 and the new q is equal to q = $49 - \binom{7}{4} = 14$ and a unit will be added to the fifth last places.

```
  2 3 4 5 6
+ 1 1 1 1 1
-----------
  2 3 4 5 6
```

Till this step, ascending sequence is $[2, 3, 4, 5, 6]$.

Because we went one step ahead in the previous stage, we continue the algorithm from column $n - m - p = 8 - 5 - 1 = 2$. According to Table 3, for q = 14 we have $\binom{5}{3} < q < \binom{6}{4}$. Since we start moving from the fourth row and third column, which equals to $\binom{5}{3}$. Then, we have $q \geq \binom{5}{3} + \binom{4}{3}$ and p=2. So, two units are added to the last four places.

```
  2 3 4 5 6
+   2 2 2 2
-----------
  2 5 6 7 8
```

Since, the new value of q is q = $14 - \left(\binom{5}{3} + \binom{4}{3}\right) = 0$, the algorithm has finished and 49th element in the sequence of dictionary order is generated.

$$B_{49} = [2,5,6,7,8]$$

It is proven for any arbitrary $q$, using combinatorial addition the whole dictionary ordered elements are produced.

Theorem 2: Using combinatorial addition, by adding arbitrary $q$ to First Member of the dictionary order, exactly $q^{th}$ element in the dictionary order for $m < n$ will be generated.

Proof: We will show this theorem using mathematical induction

First **step k = 1:**

To produce the second ascending sequence from the First Member, regarding to Combinatorial Addition, just it needs to add a unit to the first place.

$$B_0 = [1,2,3,\dots,m-1,m]$$
$$B_1 = \left[1,2,3,\dots,m-1,\underbrace{m+1}_{k=1}\right]$$

This is the second element in dictionary order. In other words, only one location was changed.

**Inductive assumption:**

Suppose using combinatorial addition to add $q$ units to First Member. Thus, the ascending sequence $1,2,3,\dots,a_{m-k+1},\underbrace{a_{m-k},a_{m-k-1},\dots,a_m}_{k}$ is produced, which is exactly the $q^{th}$ element in dictionary order. This sequence is obtained by changing at most $k$ places of First Member.

**Inductive rule:**

It should be shown that adding $q + 1$ units to First Member, the $(q+1)^{th}$ element in the dictionary order will be generated.

First case: Suppose by adding $q + 1$ to First Member, just $k$ places have changed. Regarding the inductive assumption, since only $k$ places were changed, therefore, it is exactly the $(q + 1)^{th}$ element in the dictionary order.

$$1,2,3,\dots,a_{m-k+1},\underbrace{a_{m-k},a_{m-k-1},\dots,a_m}_{k}$$

Second case: by adding q to First Member, if the ascending sequence $1,2,3,\dots,\underbrace{m-k,n-k+1,\dots,n}_{k}$ is generated, it will be impossible to increase any of the last $k$ places. Because, all places achieve to their highest possible value. According to the dictionary order, it's clear the $(q + 1)^{th}$ element is $1,2,3,\dots,\underbrace{m-k+1,m-k+2,\dots,m+1}_{k+1}$.

We will show Combinatorial Addition exactly generated the same sequence. According to Table 1, the value is equal to

$$q = \underbrace{\binom{k+n-(m+1)}{k-1} + \binom{k+n-(m+2)}{k-1} + \dots + \binom{k}{k-1}}_{n-m}$$

If we add one unit to both sides then we will have

$$q + 1 = \underbrace{\binom{k+n-(m+1)}{k-1} + \binom{k+n-(m+2)}{k-1} + \cdots + \binom{k}{k-1}}_{n-m} + 1$$

In this equation, the right side is equal to

$$\binom{k+n-(m+1)}{k-1} + \binom{k+n-(m+2)}{k-1} + \cdots + \binom{k}{k-1} + 1 = \binom{k+n-m}{k}$$

According to the above equation we have

$$q + 1 = \binom{k+n-m}{k}$$

Defined as Combinatorial Addition, the $(k + 1)^{th}$ place has increased a unit and other elements subsequently increased. So, the following ascending sequence is obtained.

$$1,2,3, \ldots, \underbrace{m - k + 1, m - k + 2, \ldots, m + 1}_{k+1}$$

This sequence is exactly $(q + 1)^{th}$ element in the dictionary order

## 5. PARALLEL ALGORITHM FOR COMPUTING RAIDC'S DETERMINANT

In this section, we present an efficient algorithm to produce the square sub matrices of definition (3) in parallel.

The algorithm is able to receive the value of $q$ and for arbitrary $m$ and $n$ produce the $q^{th}$ sequence in the dictionary order. Pseudo code for this algorithm is shown in Figure 1.

```
For i = 1 To (n - m+ 1)
        A(1, i) = i
For i = 1 To m
        A(i, 1) = 1
k = n - m+ 1
For i = 2 To m
     For j = 2 To k
        A(i, j) = A(i, j - 1) + A(i - 1, j)
j = 1
For i = 1 To m
     B(i) = i
Sum = 0, p = 0, i = k
While A(j, k) <= q
        j = j + 1
j = j - 1
i = k
     While Sum <= q
        Sum = A(j, i) + Sum
        p = p + 1
        i = i - 1
     Sum = Sum - A(j, i + 1)
     p = p - 1
     B(m - j) = B(m - j) + p
     For h = m - j To m - 1
        B(h+ 1) = B(h) + p
     q= q - Sum
     j = 1
     k = k - p
     p = 0
     Sum = 0
Wend
B(m) = B(m) + q
```

Fig 1: Pseudo code of generating arbitrary sequence

This algorithm can be implemented in various granularities. This means whatever the number of processors is further, the granularities can be smaller. And we will have larger granularities if the number of processors is less. In other words, if the number of processors is k, the number of granularities will be $\frac{\binom{n}{m}}{k}$. It means the first processor starts from $B_0$ to $B_{\frac{\binom{n}{m}}{k}-1}$ and the next portion form $B_{\frac{\binom{n}{m}}{k}}$ to $B_{2\times\frac{\binom{n}{m}}{k}-1}$ is for the second processor. In the same way, the last processor calculates $B_{(k-1)\times\frac{\binom{n}{m}}{k}}$ to $B_{\binom{n}{m}-1}$. Pseudo-code for producing ascending sequence from a specific element has been shown in figure 1.

```
for num = 1 to  (n m) / k  - 1
     B(i) = B(i) + 1
     If B(i) > n Then
        B(i - k) = B(i - k) + 1
        While B(i - k) > n - k
        k = k + 1
        B(i - k) = B(i - k) + 1
     Wend
     If (k < m) Then
```

```
    For l = i − k To m − 1
        B(l + 1) = B(l) + 1
        k = 1
    End If
    End If
end
```

Figure 1: dictionary sequence

## 6. ALGORITHM ANALYSIS DUE TO THE HARDWARE ARCHITECTURE

The proposed algorithm has the ability to run on different architectures. Parallel Random Access Machine (PRAM) is a shared memory abstract machine. In this architecture shared memory plays an important role.

On Concurrent Read Concurrent Write (CRCW) memory, the highest performance of the algorithm can be achieved. In this case, if we have $\binom{n}{m}$ processors, each processor is only run the algorithm, which is shown in Figure 1, once to obtain the corresponding square matrix with $O\big(m(n-m)\big)$ time complexity.

According to the algorithm presented in [7], if we have $m^2$ processors, the determinant of each $m \times m$ square matrix is calculated with $O(m)$. Therefore, if we have $m^2 \times \binom{n}{m}$ processors with a CRCW memory, this algorithm can calculate the determinant of $m \times n$ non-square matrix in $O(m(n-m)+m) \in O\big(m(n-m)\big)$.

If the memory is Concurrent Read Exclusive Write (CREW), the time required to sum the results of all processors in tree structure will be equal to $log\binom{n}{m}$. we know that $logm! \in \theta(mlogm)$. Thus the determinant of the non-square matrix will be calculated at $O(m(n-m)+mlogm) \in O\big(m(n-m)\big)$.

In Exclusive Read Exclusive Write (EREW) memories, there is a burden to read matrices. If enough memory is available, the matrix can be copied in a tree structure in $log\binom{n}{m}$ time complexity. Then, it will be accessible for all processors. In this case, the algorithm complexity is $O(m(n-m)+2mlogm) \in O\big(m(n-m)\big)$. Given the above description it has been shown that the time complexity of the proposed algorithm is $O(n^2)$.

It's obvious the proposed algorithm in cloud computing architecture and other architectures in which processors are connected through the network tolerates the overhead of network too. So it's time complexity will be $O(n^2 + network\_overhead)$.

## 7. CONCLUSION

Using Parallel algorithms is an effective method for reducing the time complexity. However, in most cases, increasing the number of processors does not increase productivity and just reduces the required time. But given that the cost of producing complex hardware with many processors is declining sharply, therefore the parallel algorithm can have appropriate efficiency.

On the other hand, time is an important factor in reducing the response time of real-time systems, and it plays a key role in the success of such systems. Note that, also in the machine vision, time is one of the important factors; the proposed algorithm can be very efficient and effective.

## 8. FUTURE WORK

In recent years, researchers interest in Cloud computing and distributed processing. Since the proposed algorithm can be implemented in distributed systems, implementation and computing network overhead in these systems can be considered as future researches.

With regard to applications of the determinant of matrix in image and video processing, making a proper hardware and implementing the proposed algorithm can be a suitable solution in computer vision.

There are other definition for determinant of non-square matrices, these definition can be investigated whether they can be parallelize or not and be compared with proposed algorithm in this paper.

## Authors

**Neda Abdollahi** was born in Zanjan, Iran, in 1985. She received the B.S. and M.Sc. Islamic Azad University, Zanjan, Iran, in 2008 and 2011. In 2009, she joined Bina Software Co., as a technical expert and science then she has been with Department of Electronic and Computer Engineering, Islamic Azad University of Zanjan and Payame Noor University and from 2012, she has been a Faculty member of Saeb University, Abhar, Iran. Her research interests include Multimedia Retrieval, Software Modeling, Machine Learning Methods, Object-Oriented Analysis and Design, Mobile Ad-Hoc Networks, Distributed Systems, Micro Programming and Web Design.

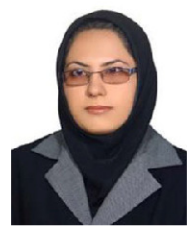

**Mohammad Jafari** was born in Zanjan, Iran, in 1977. She received the B.S. and M.Sc. Islamic Azad University, Zanjan, Iran, in 2003 and 2011. In 2001, he has started his work as CEO of Bina Software Co., Zanjan, Iran. From 2006 to 2011, he was a computer technical expert of information and planning unit of the Zanjan Broadcasting Center. And science 2009, he has been with Department of Electronic and Computer Engineering, Islamic Azad University of Zanjan and University of Applied Science and Technology and Sufi University, Zanjan, Iran. his research interests include Software Modeling, Data Base, Object-Oriented Design, Computer Networking, Mobile Distributed Systems, Programming and Web Design.

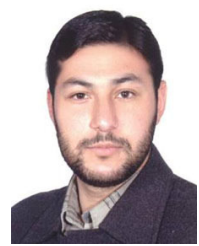

**Morteza Bayat** received M.S. degree in 2002 from Institute for Advance Studies in Basic Sciences (IASBS), Zanjan, Iran, in Applied Mathematics, and the Ph.D. degree in Differential Equations in 2008 from IASBS. Since 2008, he has worked in the Computer Engineering Department of Zanjan University as an Professor Assistant. His research interests include Matrix Computations and Differential Equations.

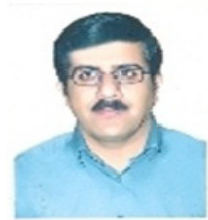

**Ali Amiri** was born in Zanjan, Iran, in 1982. He received the M.S. degree in Computer Engineering from Iran University of Science and Technology (IUST), in 2006, where he is currently pursuing Ph.D. degree in the IUST. His research interests include video segmentation, video retrieval and summarization, Matrix Computation and moving object detection and tracking.

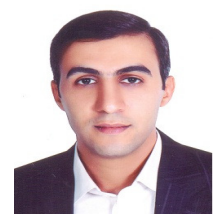

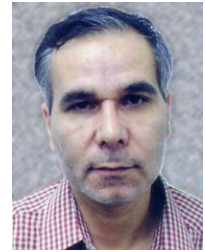

**Mahmood Fathy**, was born in Tehran, Iran, in 1959. He received the B.S. degree in Electronic Engineering from Iran University of Science and Technology (IUST), in 1985, M.Sc. degree in Microprocessor Engineering from Bradford, UK, 1988 and Ph.D. degree in Image Processing and Processor Design from UMIST, UK in 1991. Since 1992, he has been with the IUST, where he is currently Associate Professor at the Department of Computer Engineering. His research interests include Computer networks, QOS , internet Engineering, Application of image processing in Traffic, Computer Architecture for image processing, video processing applications, Panorama, Supper resolution, video classification, video retrieval and summarization.